\documentclass[conference]{IEEEtran}

\usepackage{amsmath,amssymb,amsfonts}
\usepackage{graphicx}
\usepackage{xcolor}
\usepackage{float}
\usepackage{multirow,enumitem}
\usepackage{subcaption}
\usepackage{graphicx}
\usepackage{float}
\usepackage{listings}
\usepackage{algorithmic}
\usepackage{textcomp}
\usepackage{datatool,booktabs}
\usepackage{tabularray}
\usepackage{changepage}
\usepackage{tabularx}
\usepackage{tikz}
\usetikzlibrary{shapes.geometric}
\usepackage{verbatim}

\newcommand{\circled}[1]{\tikz[baseline=(char.base)]{
    \node[shape=circle,fill=white,text=black,draw=black,inner sep=1pt] (char) {#1};}}

\definecolor{bestperf}{HTML}{FF0000} % Navy Blue
\begin{document}
\title{MARS: A Monte Carlo Tree Search-based Adaptive and Responsive Scheduler}

% \author{\IEEEauthorblockN{Michael Shell}
% \IEEEauthorblockA{School of Electrical and\\Computer Engineering\\
% Georgia Institute of Technology\\
% Atlanta, Georgia 30332--0250\\
% Email: http://www.michaelshell.org/contact.html}
% \and
% \IEEEauthorblockN{Homer Simpson}
% \IEEEauthorblockA{Twentieth Century Fox\\
% Email: homer@thesimpsons.com}
% \and
% \IEEEauthorblockN{James Kirk\\ and Montgomery Scott}
% \IEEEauthorblockA{Starfleet Academy\\
% San Francisco, California 96678--2391\\
% Telephone: (800) 555--1212\\

\author{\IEEEauthorblockN{Anonymous Authors}}

\author{
\IEEEauthorblockN{
Yash Kurkure\IEEEauthorrefmark{1},
Yihe Zhang\IEEEauthorrefmark{1},
Zhiling Lan\IEEEauthorrefmark{1}\IEEEauthorrefmark{2},
Michael E. Papka\IEEEauthorrefmark{2}\IEEEauthorrefmark{1}
}

\IEEEauthorblockA{\IEEEauthorrefmark{1}
University of Illinois Chicago, Chicago, IL, USA\\
}

\IEEEauthorblockA{\IEEEauthorrefmark{2}
Argonne National Laboratory, Lemont, IL, USA\\
\{ykurku2, yihe6, zlan, papka\}@uic.edu
}
}

\maketitle

\begin{abstract}
Modern High Performance Computing systems depend on static heuristics and manual administration for job scheduling and reservation management. Deep Reinforcement Learning (DRL) has shown promising scheduling performance but requires historical training data and fixes the optimization goal at training time, forcing operators to retrain whenever priorities shift. We introduce \textbf{MARS} (\underline{\textbf{M}}onte Carlo Tree Search-based \underline{\textbf{A}}daptive and \underline{\textbf{R}}esponsive \underline{\textbf{S}}cheduler), a training-free HPC scheduler whose optimization goal is configurable through a reward function rather than baked into a learned model. MARS uses a lightweight discrete-event simulator to explore the future consequences of scheduling decisions within a strict time budget, adapting to the configured reward at each scheduling cycle. We evaluate MARS on year-long production workloads from two systems at Argonne Leadership Computing Facility -- 4,360-node Theta and 560-node Polaris---under two reward functions: wait-time minimization (MARS-CW) and utilization maximization (MARS-CU). Unlike DRL and heuristics, which only react to the current queue or wait for backfill to find holes, MARS exploits look-ahead to proactively drain the system and plan around future reservations, packing the system to avoid the fragmentation and utilization drop that typically precede reservation windows. MARS-CW reduces tail wait time by 64\% on Theta and 43\% on Polaris over the production WFP heuristic, while MARS-CU recovers utilization in the 48 hours leading into maintenance, demonstrating that  MARS can target either objective via reward reconfiguration.
\end{abstract}

\begin{IEEEkeywords}
high-performance computing, cluster scheduling, job scheduling, Monte Carlo Tree Search, discrete event simulation, system draining, reservation-aware scheduling, adaptive scheduling
\end{IEEEkeywords}

\section{Introduction}
% Leave this empty
Modern High Performance Computing (HPC) systems process massive, heterogeneous workloads in which even minor scheduling inefficiencies can waste millions of core-hours. Despite this, production centers still predominantly rely on static heuristics such as First-Come-First-Served (FCFS) combined with EASY backfilling \cite{allcock_expatanl_2018, backfill}. These heuristics are \emph{fundamentally myopic}: they make decisions based solely on the current system state and cannot reason about how those decisions will shape the system's future. Deep Reinforcement Learning (DRL) has emerged as a promising alternative, but it depends heavily on historical training data and generalizes poorly across systems \cite{zhang_rlscheduler_2020,fan_deep_2021}.  Moreover, DRL models lack the flexibility of changing optimization goals without the need of costly retraining..

The lack of flexibility in DRL and heuristics methods is particularly evident during system-wide reservations, that require the system to \emph{drain} all active jobs before a fixed deadline. Traditional heuristics and DRL schedulers cannot automate this process because their scoring functions are oblivious to upcoming reservation boundaries. To maintain high utilization during these periods, administrators must often manually select specific large jobs that fit precisely within the remaining window~\cite{allcock_expatanl_2018, ALCFThetaPolicy}—a strategic selection that reactive heuristics and DRL cannot replicate. Effectively automating these transitions requires \emph{reservation-aware look-ahead} to observe upcoming events and proactively shape the schedule in the hours leading up to the deadline. This capability is essential not only for external maintenance but also for \emph{opportunistic draining}, where a scheduler strategically delays smaller jobs to consolidate resources for large-scale requests. Bridging this gap requires a mechanism that can weigh the long-term rewards of a coordinated drain against the immediate, reactive gains of traditional backfilling.

\begin{figure}[ht]
  \centering
  \includegraphics[width=0.68\linewidth]{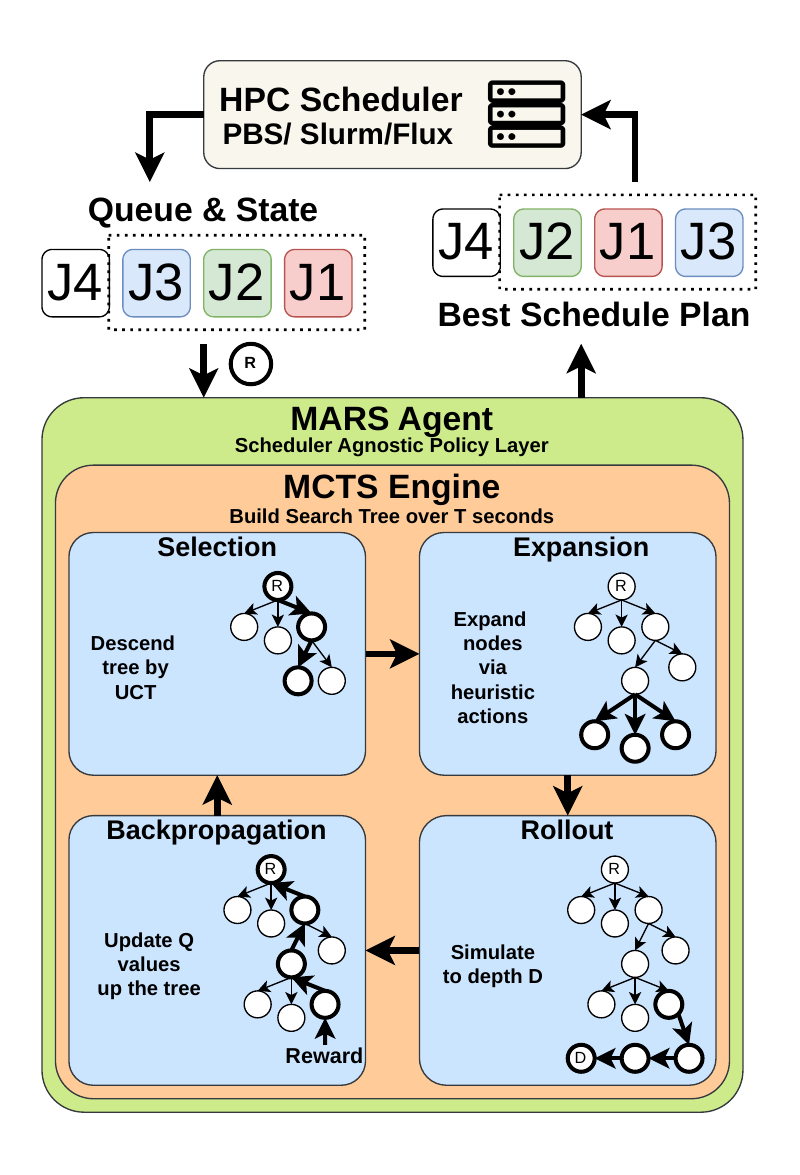}
  \caption{\small{MARS overview. At each scheduling cycle, the agent receives the current system state and waiting queue, and initializes the MCTS engine with a root node \protect\circled{R} representing the present state.  
  MCTS then iterates over four stages: Selection, Expansion, Rollout, and Backpropagation for a fixed budget of $T$ seconds. When the budget expires, the action leading to the most-visited child of the root is returned as the schedule plan and applied to the production scheduler by adjusting job priorities accordingly.}}
  \label{fig:overview}
\end{figure}

To bridge this gap between rigid heuristics and inflexible DRL models, we introduce \textbf{MARS} (\underline{\textbf{M}}onte Carlo Tree Search-based \underline{\textbf{A}}daptive and \underline{\textbf{R}}esponsive \underline{\textbf{S}}cheduler). Monte Carlo Tree Search (MCTS) is a heuristic search algorithm that achieved widespread recognition through AlphaGo~\cite{silver_mastering_2017, silver_mastering_chess_shogi_2017} for sequential decision-making by incrementally constructing a search tree where nodes represent states and edges represent actions. We formulate HPC scheduling as \emph{an online planning problem solved via MCTS}, where the optimization goal is encoded as a configurable reward function rather than fixed at training time. This makes MARS \emph{adaptive}: the search shapes its decisions around the configured goal without any retraining. By simulating the consequences of reordering the queue within a fixed time budget $T$, MARS is also \emph{responsive}: it reacts to the live system state at each scheduling cycle and delivers a decision when the budget expires. This look-ahead capability enables MARS to proactively create scheduling holes through system draining, rather than waiting for backfill to find holes reactively. Figure~\ref{fig:overview} illustrates this closed-loop, cycling through Selection, Expansion, Rollout, and Backpropagation stages of MCTS before dispatching the best action.

Transitioning MCTS from board games to the dynamic state space of HPC scheduling introduces \emph{two major challenges}: (i) an extremely large branching factor induced by combinatorial job-to-node assignments, and (ii) strict real-time constraints that require scheduling decisions within seconds. We address these challenges through the following techniques:
\begin{enumerate}
    \item \textbf{MDP formulation with MCTS:} 
    We model the HPC scheduling problem as a Markov Decision Process (MDP), enabling principled sequential decision-making under uncertainty in a structured state-action space.
%    We formulate the HPC scheduling process as a Markov Decision Process (MDP).

    \item \textbf{Adaptive goal configuration:} 
    MARS utilizes configurable reward functions to optimize user-level (wait time) or system-level (utilization) objectives, adapting its search behavior at each scheduling cycle without retraining.

    \item \textbf{Responsive scheduling under time constraints:} 
    We introduce tree pruning, heuristic branching, and root parallelization that enable MARS to produce scheduling decisions within a 15-second scheduling budget.
\end{enumerate}

We evaluate our design on year-long production workloads from two production HPC systems at Argonne Leadership Computing Facility (ALCF) of contrasting scale: Polaris (560 nodes, 2024) and Theta (4{,}360 nodes, 2021), demonstrating that MARS scales across an order-of-magnitude difference in system size. We analyze wait time and utilization, and further break down the results by uptime and downtime regions. Our results show that MARS adapts its scheduling behavior to the configured reward: rather than waiting for backfill to find scheduling holes reactively, MARS-CW (wait time) proactively creates holes by draining the system to reduce job wait times, while MARS-CU (utilization) applies the same look-ahead to improve utilization in the 48 hours leading into maintenance windows.

The rest of the paper is organized as follows. Section~II reviews HPC scheduling. Section~III formulates the problem as an MDP solved via MCTS. Section~IV presents the techniques. Sections~V - VI cover the experimental results on Polaris and Theta. Section~VII surveys related work, and Section~VIII concludes.

\section{Background}
\textbf{Scheduling in HPC.} Scheduling in HPC systems is handled by job schedulers such as PBSPro \cite{AltairPBSPro}, Slurm \cite{SlurmDocs}, or Flux \cite{FluxFramework}. Users submit jobs using the scheduler's interface and provide the resource requirements and walltime (duration of job). In a production HPC system, the smallest schedulable unit is a node that may contain multiple CPUs and GPUs. This is because HPC centers prefer giving researchers dedicated bare metal resources to avoid interference from other users on the system. Thus, each job is characterized by the number of nodes (job size) it requires and the walltime. The order in which the submitted jobs run of the system is dictated by the scheduling policy.

\textbf{Scheduling Policy.}  In academic work, the scheduling policy is generally viewed as the rule that determines the order in which the submitted jobs run on the system. In practice, this is much more complex. HPC facilities must perform system maintenance, meet allocation award goals of funding programs, and reserve the whole or parts of the system for special requests \cite{allcock_expatanl_2018, ALCFResvPolicy, OLCFResvPolicy}. Such nuances have been ignored by the state-of-the-art scheduling, especially AI-based intelligent policies \cite{fan_deep_2021, zhang_rlscheduler_2020}. We fill in this gap by proposing MARS, which is intelligent and can easily adapt to such goals. While \S\ref{sec:related_work} expands more on prior work, Table \ref{tab:scheduling-comparison} provides an overview of the comparison.

\begin{table}[h]
\centering
\small % Slights reduces font size
\setlength{\tabcolsep}{3pt} % Reduces horizontal padding (default is 6pt)
\begin{tabular}{|p{3.5cm}|c|c|c|c|} 
 \hline
 \textbf{Features} & \textbf{Heuristics} & \textbf{Optimization} & \textbf{DRL} & \textbf{MCTS} \\
 \hline
 Training Required & No & No & Yes & No \\
 \hline
 Cross System Stability & Yes & Yes & No & Yes \\
 \hline
 Goal Flexibility & No & Yes & No & Yes \\
 \hline
 Machine 
 Reservations & No & No & No & Yes \\
 \hline
 System Draining & No & No & No & Yes \\
 \hline
\end{tabular}
\caption{\small{Comparison of scheduling methods.}}
\label{tab:scheduling-comparison}
\end{table}

\section{Problem Formulation}
\label{sec:problem_formulation}

We formulate HPC job scheduling as a sequential MDP and solve it online via MCTS. MARS acts as a scheduler-agnostic policy layer that augments existing production schedulers with an MCTS agent, performing look-ahead planning through a Discrete Event Simulation (DES) of the system. This framework allows the agent to evaluate the long-term impacts of counterfactual actions at each decision point — without disturbing the production system. The remainder of this section details the DES model and its MDP formulation, then describes how MDP states are embedded within the four MCTS stages.

\subsection{Scheduling Model} 
\label{des}

The scheduling model serves two roles: it represents the states stored in the search tree, and it encodes the dynamics that govern how the system evolves when an action is applied. We model HPC scheduling using DES, in which system time advances between discrete events. This paradigm is widely used in HPC research, for strategy evaluation \cite{rodrigo2018scsf, dutot2017batsim, klusavcek2020alea, galleguillos2020accasim} and as a training environment for DRL-based models \cite{zhang_rlscheduler_2020, fan_deep_2021, yang_deep_2022, zhang_schedinspector_2022}. \emph{MARS adopts the same paradigm but uses it online}: rather than evaluating a fixed policy in simulation, the DES is invoked at every scheduling cycle as a forward model for MCTS rollouts.

The simulator processes events chronologically, with each event responsible for updating the system state and scheduling future events. We consider the following event types:

\begin{itemize}
    \item \textit{Job Submit/Run/End}: Standard job lifecycle transitions.
    \item \textit{Scheduling Cycle}: Invokes the scheduling mechanism to dispatch jobs.
    \item \textit{Announce Downtime}: Occurs 48 hours prior to scheduled maintenance. Beyond this point, the scheduler prevents the execution of any job whose walltime overlaps with the maintenance window.
    \item \textit{Scheduled Downtime Begin/End}: Marks the start and completion of planned maintenance.
    \item \textit{Unscheduled Downtime Begin/End}: Represents emergency failures. In these instances, all running jobs are terminated and returned to the queue for resubmission.
\end{itemize}

The simulation follows a reactive loop: a \textit{Job Submit} event triggers a \textit{Scheduling Cycle}; if resources permit, the cycle inserts \textit{Job Run} events, which in turn generate \textit{Job End} events based on the actual runtime. Every job completion (\textit{Job End}) triggers a new \textit{Scheduling Cycle} to fill the newly available resources. The core capability of MARS lies in its ability to clone the system state at any \textit{Scheduling Cycle} and replay it under alternative scheduling decisions, enabling the agent to evaluate the long-term consequences of different actions.

\subsection{Scheduling Window}
The \emph{scheduling window} is a technique widely adopted in prior scheduling work \cite{yang_deep_2022, yang-sc13, zhang_rlscheduler_2020} that constrains the agent's action space to a tractable subset. The window contains the first \textit{w} jobs at the head of the waiting queue. Within this window, the agent reorders jobs by priority. The scheduler then processes the reordered queue sequentially until encountering a \textbf{top job} that cannot be scheduled due to resource constraints, followed by an EASY backfilling procedure that attempts to utilize remaining nodes without delaying the top job based on its predicted start time.

MARS is unique in its ability to leverage the integrated simulator to evaluate alternative orderings during rollouts. Rather than committing to a single heuristic, MARS explores the decision space by testing various reorderings within its rollouts, thereby discovering scheduling policies that outperform traditional fixed-priority approaches.

\subsection{Markov Decision Process}
\label{sec:mdp}
The above scheduling process can be represented as a Markov Decision Process (MDP) defined by the tuple $(\mathcal{S}, \mathcal{A}, \mathcal{T}, \mathcal{R}, \gamma)$:

\begin{itemize}
    \item \textit{State} $s_t \in \mathcal{S}$: A complete snapshot of the simulator at a \textit{Scheduling Cycle} event, comprising the queue of waiting jobs, the set of running jobs with their completion times, the number of available nodes, the current simulation time, and the event queue of future events. This snapshot is taken just before the scheduler has run and inserted any new \textit{Job Run} events into the event queue.

    \item \textit{Action} $a_t \in \mathcal{A}$: A reordering of the \emph{window} which is the first $w$ jobs at the head of the queue of waiting jobs.

    \item \textit{Transition} $\mathcal{T}(s_{t+1} | s_t, a_t)$: The action reorders the first $w$ jobs in the queue, after which the scheduler runs. The transition then fast-forwards through all non-\textit{Scheduling Cycle} events until the next \textit{Scheduling Cycle} event is reached.

    \item \textit{Reward} $\mathcal{R}(s_{t+1})$: An immediate reward assigned to the successor state reached after applying action $a_t$ in state $s_t$. It is computed from the set of jobs dispatched during the transition from $s_t$ to $s_{t+1}$.

    \item \textit{Discount factor} $\gamma \in [0, 1]$: The discount factor is multiplied with the reward depending on how far ahead it is in the future. A higher discount factor gives more emphasis to future rewards than the present, making MCTS less greedy.
\end{itemize}

A sequence of such states and actions forms a Markov decision process, which follows the Markov property that the future states only depend on the current state and not the history of actions and states preceding it:

\begin{equation}
    s_{\text{0}} \xrightarrow{a_0} s_1 \xrightarrow{a_1} \cdots \xrightarrow{a_k} s_{\text{k}}
\end{equation}

\subsection{Monte Carlo Tree Search (MCTS)}
\label{sec:mcts}

The core function of MCTS is \emph{to build a search tree that explores the space of action sequences, balancing exploration of untried actions against exploitation of high-reward ones}. The MDP formulation maps onto this tree directly: each state $s_t \in \mathcal{S}$ is represented as a node, and each action $a_t \in \mathcal{A}$ is an edge connecting a parent state to the successor state produced by the transition. Repeated application yields a tree rooted at the current system state, with each path from root to leaf representing one trajectory of look-ahead.

The algorithm is invoked at each \emph{Scheduling Cycle} event where the waiting queue contains more than one job. First, the root node of the tree is initiated using the current state of the system. Then MCTS iteratively builds a tree of system states and actions using the four stages of selection, expansion, rollout, and backpropagation. Finally, the algorithm returns the best window ordering it found from the current state, which is used by the scheduler to run new jobs. 

\subsubsection{Selection}
\label{sec:selection}
Starting from the root node, the algorithm descends the tree by repeatedly selecting the child with the highest UCT score. The UCT (Upper Confidence Bound applied to Trees) score is the formula used in MCTS to balance exploitation (favoring moves that have performed well) against exploration (trying less-visited moves). It is defined as:
\begin{equation}
    \text{UCT}(s_t) = Q(s_t) + c \cdot \sqrt{\frac{2 \ln N}{n_t}}
\end{equation}
where $Q(s_t)$ is the expected reward beyond state $s_t$, $n_t$ is number of times node was visited, $N$ is the parent's visit count, and $c$ is the exploration constant. The first term is greedy in terms of selecting nodes with the highest average reward, whereas the second term grows inversely with $n_t$, promoting exploration of less visited nodes in the tree. A larger $c$ would thus promote exploration and reduce greedy behavior. The selection stage continues until it reaches a leaf of the tree. The leaf node is then passed to the next stage.

\subsubsection{Expansion}
\label{sec:expansion}
The expansion stage takes the leaf node and instantiates a child node by 
(1) copying the leaf node's simulator state, (2) reordering the queue according to the child's action, and (3) stepping the simulation forward until the next Scheduling Cycle event. This is represented by the transition $\mathcal{T}(s_{t+1} | s_t, a_{t})$. The \emph{outcome} of this action $a_{t}$ is the set of jobs $o_{t}$ that go from waiting to running, which is recorded to compute the immediate reward $\mathcal{R}(s_{t+1})$ of the node's state. When all possible actions from the leaf lead to a child node, it stops being a leaf node, and its simulator state is released from memory to conserve resources, as it won't be needed for future expansions or rollouts.

\subsubsection{Rollout}
\label{sec:rollout}
From the newly expanded leaf, the simulation is run forward with actions of random orderings of the window until termination or a predefined maximum depth for the tree is reached. The process so far can be summed up as follows:

\begin{equation}
\underbrace{s_{\text{root}} \xrightarrow{a_0} s_1 \xrightarrow{a_1} \cdots \xrightarrow{a_{k-1}} s_{\text{leaf}}}_{\text{selection}} \underbrace{\xrightarrow{a_{\text{k}}} s_{k+1}}_{\text{expansion}} \underbrace{\xrightarrow{a_{\text{rand}}} \cdots \xrightarrow{a_{\text{rand}}} s_{D}}_{\text{rollout}}
\end{equation}

where $D$ is the predefined tree depth which controls how far in the future MCTS explores.

The reward from the rollout phase is calculated by summing up the immediate rewards of state $s_{k+1}$ to state $s_{D}$ as follows:
\begin{equation}
    G_{\text{rollout}} = (1 - \gamma)\sum_{t=k+1}^{D} \gamma^{t -(k+1)} \cdot \mathcal{R}(s_t)
\end{equation}

The $(1 - \gamma)$ normalizes the reward between $[0,1]$, which is a common approach in MCTS to ensure the UCT in the selection phase is properly calibrated.

\subsubsection{Backpropagation} 
\label{sec:backpropagation}
The return $G_{rollout}$ is then propagated from the leaf back to the root. For each node $t$ representing state $s_t$, $n_t$ is incremented and the expected value $Q(s_t)$ is updated using the rollout reward and immediate reward $\mathcal{R}(s_t)$ as follows:
\begin{equation}
    Q(s_t) \gets Q(s_t) + \frac{1}{n_t} \left[ \Big( (1 - \gamma)\mathcal{R}(s_t) + \gamma \cdot G_{\text{rollout}} \Big) - Q(s_t) \right]
\end{equation}

In simple words, $Q(s_t)$ is the average rollout reward from the node $s_t$ and is used in the selection phase for the UCT equation, which guides the search.

\section{MARS: Scaling MCTS for HPC Scheduling}
\label{sec:implementation}

Section~\ref{sec:problem_formulation} formulated HPC scheduling as an MDP, enabling MCTS to drive what-if exploration of scheduling decisions. This section presents the key techniques MARS uses to make MCTS-driven scheduling viable in production: heuristic branching to tame the action space, tree pruning to control search complexity, configurable reward to support runtime goal selection, and the resulting backfilling, reservation, and draining behaviors that emerge from these mechanisms.

\subsection{Heuristic Branching}
\label{sec:heuristic_branching}

\begin{table*}[t]
\centering
\small
\setlength{\tabcolsep}{6pt}
\begin{tabularx}{\textwidth}{ll>{\raggedright\arraybackslash}X}
\toprule
\textbf{Policy} & \textbf{Scoring Function} & \textbf{Description} \\
\midrule
FCFS & $s_j$ & First Come First Served\\
LCFS & $-s_j$ & Last Come First Served\\
SJF  & $r_j$ & Shortest Job First\\
LJF  & $-r_j$ & Longest Job First \\
SRF  & $n_j$ & Smallest Resource First\\
LRF  & $-n_j$ & Largest Resource First\\
SCF  & $r_j \cdot n_j$ & Shortest Core Hours First \\
LCF  & $-(r_j \cdot n_j)$ & Largest Core Hours First \\
FCSJ & $-w_j / r_j$ & First Come Shortest Job\\
WFP1 & $-(w_j / r_j) \cdot n_j$ & Favors old/short jobs, avoiding large job starvation \cite{tang-cluster09}.\\
WFP3 & $-(w_j / r_j)^3 \cdot n_j$ & Favors old/short jobs, avoiding large job starvation more \cite{tang-cluster09}.\\
FAT  & $-(w_j / r_j) \cdot (n_j / N)^3$ & Favors large jobs, then old/short jobs \cite{tang-cluster09}.\\
UNICEF & $-w_j / (\log_2(n_j / N) \cdot r_j)$ & Provides fast turnaround for small jobs \cite{tang-cluster09}. \\
F1   & $\log_{10}(r_j) \cdot n_j + 870 \cdot \log_{10}(s_j)$ & Favors early submissions, then small jobs; ranks mainly by node count \cite{f1_f4_ml}.\\
F2   & $\sqrt{r_j} \cdot n_j + 2.56 \times 10^4 \cdot \log_{10}(s_j)$ & Favors early submissions, then small jobs; node count dominates, walltime secondary \cite{f1_f4_ml}.\\
F3   & $r_j \cdot n_j + 6.86 \times 10^6 \cdot \log_{10}(s_j)$ & Favors early submissions, then small core-hours; penalizes walltime and nodes equally \cite{f1_f4_ml}.\\
F4   & $r_j \cdot \sqrt{n_j} + 5.30 \times 10^5 \cdot \log_{10}(s_j)$ & Favors early submissions, then small jobs; penalizes walltime over node count \cite{f1_f4_ml}.\\
\bottomrule
\end{tabularx}
\caption{\small{Heuristic policies and their respective scoring functions. Jobs are sorted in ascending order of scores calculated by the scoring functions with inputs: $r_j$ (walltime), $n_j$ (nodes), $s_j$ (submit time), $w_j$ (wait time), and $N$ (total system size).}}
\label{tab:heuristics-policies}
\end{table*}

We define each scheduling action $a_t$ as a specific ordering of jobs within a window $w$ at the head of the waiting queue. While a window of size $w$ theoretically allows for $w!$ possible permutations, this factorial growth becomes computationally prohibitive for the $w > 20$ values typically required for significant performance gains \cite{fan_deep_2021, fan_scheduling_2019, zhang_rlscheduler_2020, yang-sc13}. However, resource constraints often cause different orderings to result in the same set of jobs to be scheduled, allowing the search space to be effectively trimmed using heuristics. Following this intuition, we generate multiple queue orderings based on the 17 heuristics detailed in Table \ref{tab:heuristics-policies}, sorting the $w$ jobs at the head of the queue according to the scores produced by each heuristic function.

\begin{figure} 
    \centering
    \includegraphics[width=0.6\columnwidth]{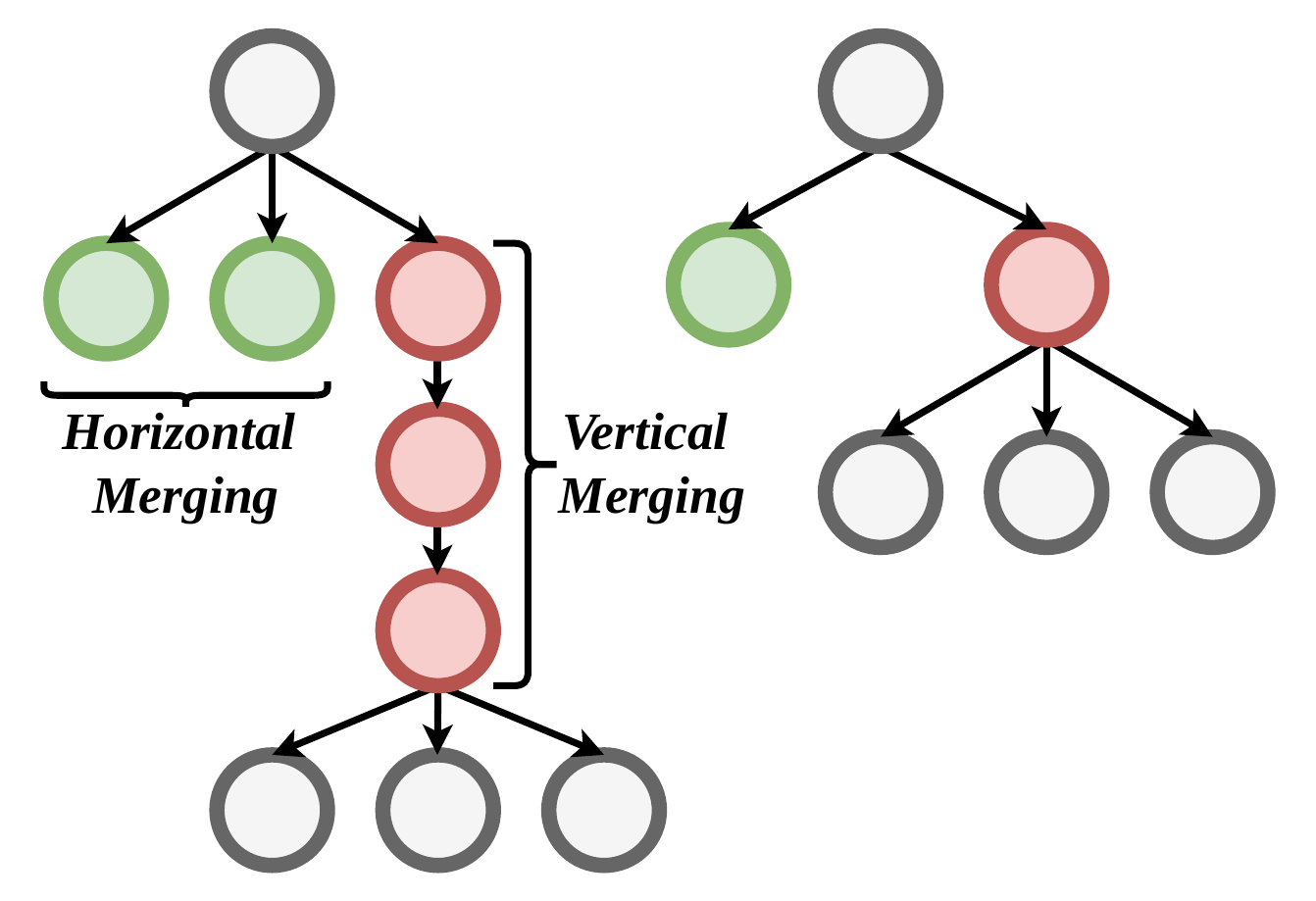}
    \caption{\small{Tree pruning via node merging. \textit{Horizontal merging} collapses sibling branches whose actions produce the same scheduling outcome; \textit{Vertical merging}  collapses consecutive scheduling cycles that yield empty outcomes (no jobs dispatched) into a single node.}}
    \label{fig:node-merge}
\end{figure}

While branching on heuristics limits the number of possible orderings, it does not determine the optimal value for $w$. We analyzed window sizes increasing by powers of two in $[2, 1024]$ across year-long workloads from Polaris (2024) and Theta (2021), finding that heuristic behavior transfers reliably across systems but the optimal window size is system-dependent. To account for these system-specific variations, we designed MARS to encompass the behavior of all 16 non-FCFS heuristics across ten window sizes ($2^1$ to $2^{10}$). By integrating these combinations with the standard FCFS baseline, MARS operates with a total branching factor of:
$16 \text{ Heuristics} \times 10 \text{ Window Sizes} + 1 \text{ FCFS} = 161$. This comprehensive action space enables the MCTS agent to dynamically adjust the trade-offs between various heuristic priorities and window sizes based on the current state of the system. We employ this branching strategy for both expansions and random sampling during Rollouts.

\subsection{Tree Pruning}
\label{sec:tree_pruning}

Due to resource constraints, we found that it is very common for multiple actions $a_t$ to lead to the same scheduling outcome $o_t$ or, multiple scheduling cycles can produce an empty outcome $o_t$. For such instances, we merge multiple nodes into a single node as shown in Figure \ref{fig:node-merge} during the expansion stage. This provides a lower bound on the branching factor of 2. As for the depth of the tree, we manually set the maximum depth to $D$ for both expansions and rollouts. Overall, the branching factor of the tree is always between $[2,161]$ and the depth does not exceed a predefined value of $D$.

\subsection{Configurable, Training-Free Reward}
\label{sec:rewards}

HPC scheduling metrics fall into two broad categories: \emph{system-level} and \emph{user-level}. The most widely reported metrics in each category are system utilization and average job wait time, respectively. To demonstrate MARS's reward-configurability, we introduce two reward functions, each targeting one of these metrics:

\subsubsection{Cumulative Wait (CW)} Wait time is a user-level metric, and we focus on optimizing for the average wait time $\overline{W_k}$ by keeping track of the jobs scheduled from the root node to some state $s_k$ in the tree. Then the immediate reward for the action is calculated as:

\begin{equation}
    \mathcal{R}_{\text{CW}}(s_k) = \frac{1}{1 + \overline{W_k}}
\end{equation}

\subsubsection{Cumulative Utilization (CU)} System admins care about system utilization, which is the total useful node hours to the available node hours during system uptime. For each action, we calculate the cumulative utilization $U_k$ from the root node to some state $s_k$ in the tree as:
\begin{equation}
    \mathcal{R}_{\text{CU}}(s_k) = U_k
\end{equation}

\subsection{Backfill-Aware Action}
\begin{figure}[ht]
  \centering
  \includegraphics[width=0.8\linewidth]{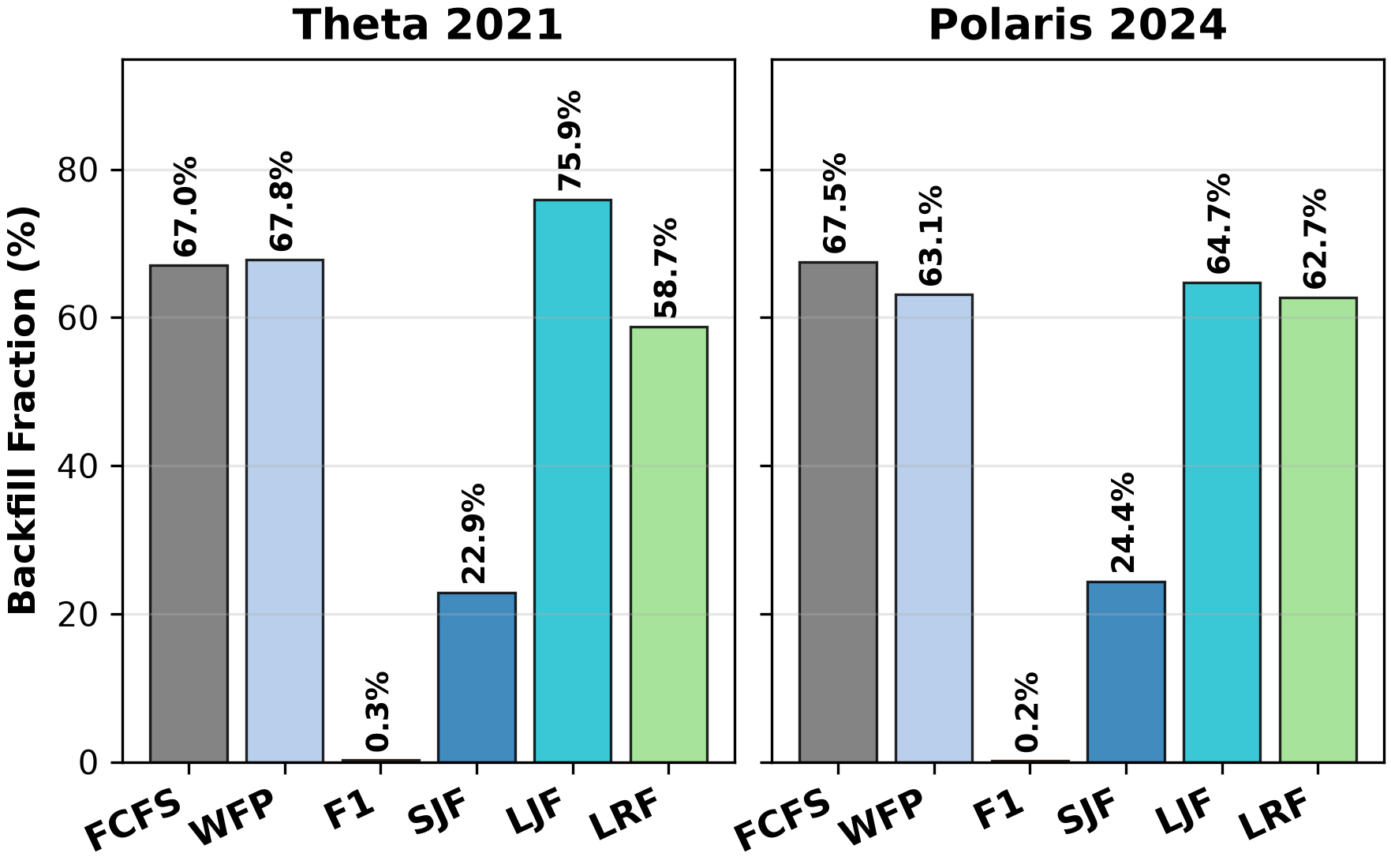}
  \caption{\small{Fraction of dispatched jobs scheduled via EASY backfilling under representative heuristics on Theta (2021) and Polaris (2024).
Backfilling usage varies by nearly two orders of magnitude across heuristics.
}}
  \label{fig:backfill_dist}
\end{figure}

EASY backfilling is the standard mechanism HPC schedulers use to fill resource holes and maximize system utilization. When paired with First-Come-First-Served (FCFS), it preserves fairness by forbidding any back-filled job from delaying the top job--the job at the head of the queue currently blocked by resource constraints. Under heuristic-based ordering, however, the top job is no longer determined by arrival time but by the utility score in Table~\ref{tab:heuristics-policies}, which fundamentally changes which jobs backfilling protects and, consequently, how often it fires.

Figure~\ref{fig:backfill_dist} shows backfill shares varying from $<$1\% (F1) to over 70\% (WFP/FCFS/LJF). This disparity reflects how heuristics manage candidate availability: F1 exhausts the pool by prioritizing small-footprint jobs, while SJF’s node-count indifference allows large, short jobs to head the queue. These blockers trap smaller jobs behind them, providing a steady supply of candidates that easily fit into backfill windows.

MARS adopts \emph{backfill-aware action semantics}: rather than treating EASY backfill as a secondary, post-policy step, the agent folds it directly into the action's outcome. During MCTS rollouts, a selected heuristic ordering and its associated backfill logic are applied in a single transition. Within the MDP, an action $a_t$ represents the choice of a window ordering under a specific heuristic at a particular $w$, while the outcome $o_t$ encompasses all jobs dispatched by both the primary ordering and the resulting backfill.

\subsection{Reservation-Aware Look-ahead}
\begin{figure}[ht]
  \centering
  \includegraphics[width=0.8\linewidth]{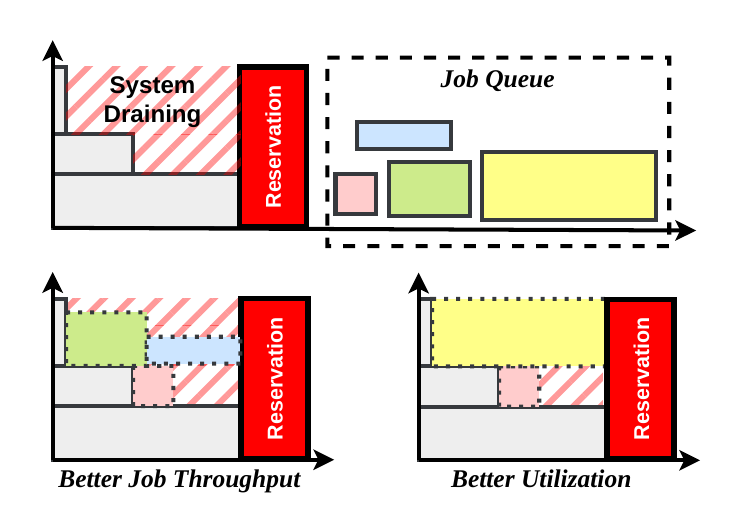}
  \caption{\small{System-wide reservation. (Top) system state before a drain, where running jobs (left) must complete and queued jobs (right) vary in size and walltime. 
  Static heuristics commit to a fixed policy and cannot distinguish between two feasible plans: \emph{Better Job Throughput} (bottom left), which packs many small jobs to reduce wait time at the cost of fragmentation, and \emph{Better Utilization} (bottom right), which schedules a single large job aligned with the reservation boundary, sacrificing throughput. MARS selects between these regimes based on the configured reward.}}

  \label{fig:reserve_example}
\end{figure}

System-wide reservations are commonly used at production HPC sites for maintenance or specialized allocations~\cite{ALCFResvPolicy, OLCFResvPolicy}. They require the scheduler to complete all active jobs before the reservation window begins, effectively forcing a system-wide drain.
As illustrated in Figure \ref{fig:reserve_example}, MARS must schedule jobs based on the specific optimization goal during this transition. The figure highlights two distinct scenarios: The left, where throughput is improved by packing the drain period with many small jobs, and the right, which prioritizes utilization by scheduling larger requests. 

% With "reservation-aware look-ahead"
Heuristics cannot navigate this trade-off because their scoring functions are fixed and oblivious to the upcoming reservation boundary. MARS, by contrast, exhibits \emph{reservation-aware look-ahead}: it observes the reservation event in the simulator's event queue and shapes scheduling decisions in the 48 hours leading up to it according to the configured reward.

\subsection{Opportunistic Draining}
\begin{figure}[ht]
  \centering
  \includegraphics[width=0.8\linewidth]{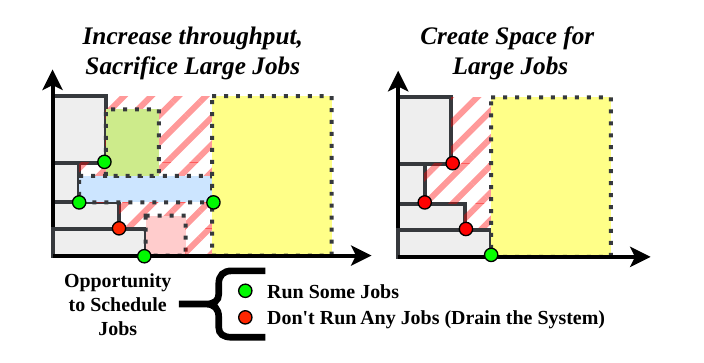}
  \caption{\small{Two schedules for the same arrival sequence. Each circle marks a scheduling cycle triggered by a job completion: green cycles dispatch new jobs, red cycles dispatch nothing (an intentional drain). \emph{Left:} greedily dispatching small jobs improves throughput but indefinitely delays the large yellow job. \emph{Right:} skipping three consecutive cycles deliberately holds resources idle until the fourth cycle, when sufficient capacity becomes available to admit the large job. MARS may drain even without an upcoming reservation to maximize the configured reward over the look-ahead horizon.}}
  \label{fig:drain_example}
\end{figure}

The reservation case forces a drain because of an externally imposed deadline. MARS generalizes this idea: at any scheduling cycle, the agent can choose to drain even without an upcoming reservation, simply because doing so maximizes the configured reward over the look-ahead horizon (i.e., \emph{opportunistic draining}).

At each scheduling cycle, MARS utilizes MCTS to select an outcome set ($o_t$) derived from a specific heuristic action ($a_t$). Because heuristics block on the first job that cannot be scheduled, $o_t$ may remain empty despite available resources, a state MARS leverages for \textit{system draining}. Figure \ref{fig:drain_example} illustrates this process: filled circles represent scheduling cycles triggered by completing jobs, where MARS can select a heuristic that either starts a new job (green circle) or starts no jobs (red circle). The left shows a situation that schedules multiple small jobs but delays the larger request. On the right, nothing is scheduled intentionally for three cycles, allowing the large job to start at the fourth. Our experiments demonstrate that MARS effectively reduces wait times for large-scale jobs by adaptively selecting these drain-inducing heuristics. This mechanism is similar to draining for reservations, except MARS can intelligently decide when to use it.

\subsection{Parallelization}

\begin{figure}[ht]
  \centering
  \includegraphics[width=0.8\linewidth]{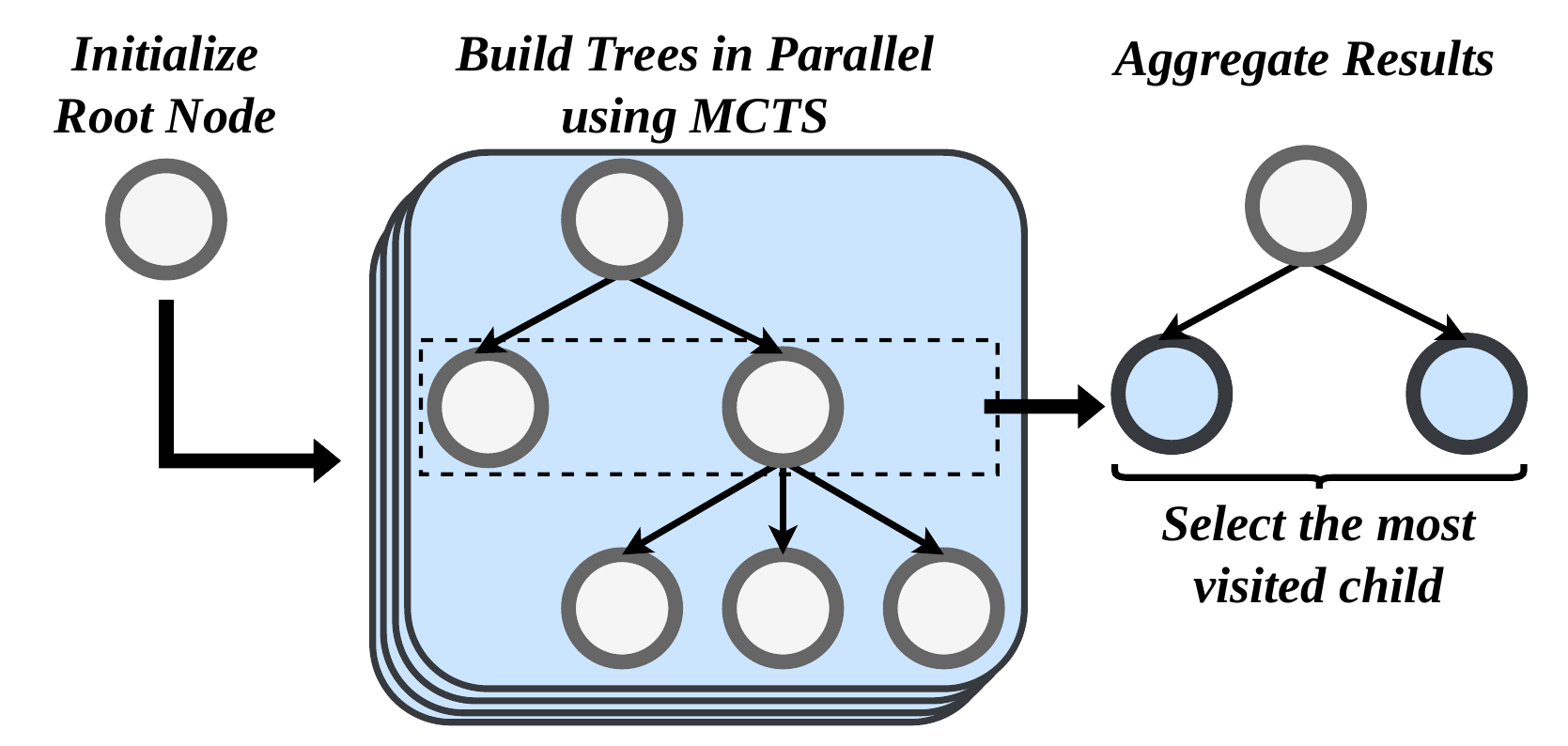}
  \caption{\small{Parallelizing MCTS using root parallelization. The root node is initialized to represent the current state of the system (left). From this root, P independent MCTS instances are executed in parallel, where P is the number of available processes, with each process constructing its own search tree (middle). After a time limit of T, the search terminates, and the root's children from all P trees are aggregated. The action corresponding to the most visited child across the aggregated trees is then selected and returned.}}

  \label{fig:mcts_parallel}
\end{figure}

The performance of MCTS is inherently time-dependent: in general, the longer the search is allowed to run, the higher the quality of the resulting decisions. In HPC scheduling, however, decisions must be made within the short interval between scheduling cycles, which has been reported to be on the order of 15 seconds \cite{yang_deep_2022, yang-sc13, tang-cluster09}. To operate within this strict time budget, MARS employs root parallelization \cite{parallel1}, as illustrated in Figure~\ref{fig:mcts_parallel}. Starting from a common root node, P independent MCTS instances are executed concurrently, where P is the number of available processes. This approach is embarrassingly parallel because no communication is required between processes during the search. Once the allotted time expires, each worker returns the visit counts of the root's children, which are then aggregated to determine the final action by selecting the most frequently visited child.

\subsection{MARS Implementation}
MARS is designed to be scheduler-agnostic, allowing it to augment existing production schedulers such as PBS or Slurm. The system is built upon the open-source discrete event simulation tool CQSim~\cite{CQSimMaster}, within which we have implemented our MARS agent. CQSim has been extensively validated in prior optimization research \cite{yang-sc13, fan_scheduling_2019} and utilized in several studies involving Deep Reinforcement Learning models \cite{fan_deep_2021, li_interpretable_2023}. While the original version of CQSim is implemented in Python, we developed a C++ implementation of the same architecture to fully exploit shared memory parallelism and achieve the computational speed necessary for complex search-based scheduling.

\section{Evaluation Methodology}

\subsection{Workload}

 \begin{figure} 
    \centering
    \includegraphics[width=0.8\columnwidth]{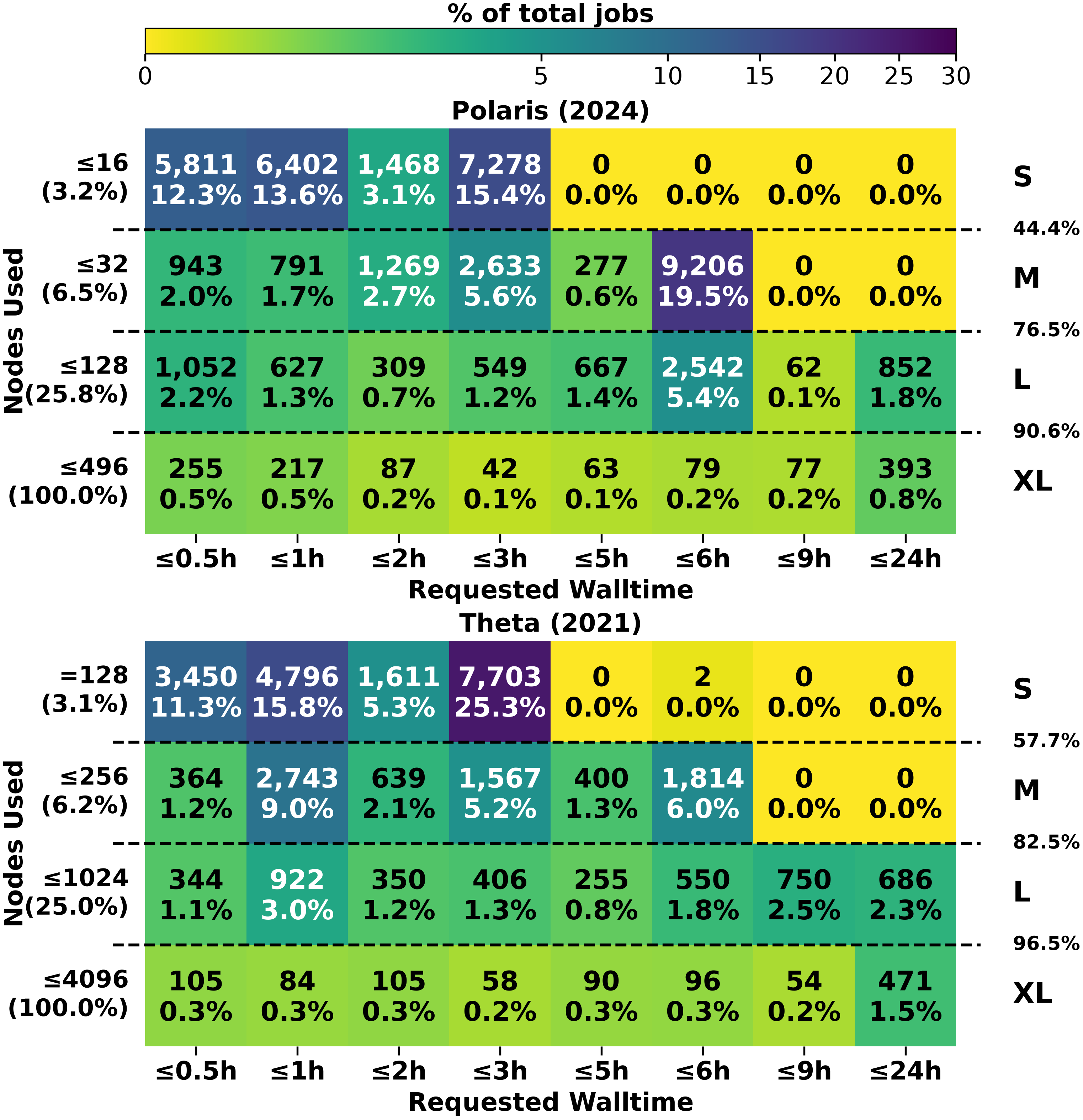}
    \caption{\small{Job distribution across the Polaris and Theta workloads, binned by node count (rows) and requested walltime (columns). Each cell shows the absolute job count and its percentage; row labels include the cumulative percentage of jobs at or below that size class. Jobs are grouped into four size classes (S, M, L, and XL) with class boundaries marked by dashed horizontal lines and cumulative shares annotated on the right. }}
    \label{fig:heatmap_workload}
\end{figure}

Our evaluations utilize two distinct production workloads from the Argonne Leadership Computing Facility (ALCF): Theta, which includes 30,415 jobs from 2021, and Polaris, which consists of 47,184 jobs from 2024. To account for nodes dynamically reserved for debugging or special requests, we simplify the system sizes to reflect the maximum job limits allowed in the production queues, resulting in 4,096 nodes for Theta and 496 nodes for Polaris. As illustrated in Figure \ref{fig:heatmap_workload}, jobs are categorized into S, M, L, and XL sizes based on their system footprint, maintaining consistent occupancy ratios across both architectures.

Both systems underwent multiple scheduled and unscheduled maintenance events. We identified these downtime regions by cross-referencing job logs with the ALCF public data portal \cite{ALCFData}, allowing us to break down our analysis into specific periods of system availability. During the recorded periods, Theta experienced 28 system-wide outages, including six unscheduled emergencies, while Polaris recorded 11 outages, two of which were unscheduled.

The workload logs provide both the user-specified walltime and the recorded actual runtime. In a live environment, the true runtime remains unknown until a job completes; therefore, the MARS planning and prediction logic relies strictly on walltimes for all decision-making. In other words, to ensure our evaluations remain grounded in reality, we replay each job using its recorded actual runtime, which allows us to faithfully reproduce real-world system behavior while adhering to the information constraints present at the time of scheduling.

\subsection{Evaluation Metrics}

\begin{itemize}
    \item \textit{Wait Time}: Wait Time is a user-level metric defined as the duration between a job's submission and its start time, which we analyze both across the total workload and categorized by specific job sizes.
    \item \textit{System Utilization}: Measures the ratio of consumed node-hours to total available capacity during system uptime. In addition to assessing utilization for each uptime period, we specifically analyze the 48-hour windows leading up to maintenance events, as these intervals are frequently characterized by low utilization due to system draining and resource fragmentation.
\end{itemize}
Both metrics are reported using box plots: boxes span the 25\textsuperscript{th} to 75\textsuperscript{th} percentiles ($P_{25}$--$P_{75}$), with the median marked inside the interquartile range ($\text{IQR}$). Whiskers extend to $\pm 1.5 \times \text{IQR}$ from the box edges, and a red dotted line marks the mean.
    
 \subsection{Baselines}

 \begin{figure} 
    \centering
    \includegraphics[width=0.8\columnwidth]{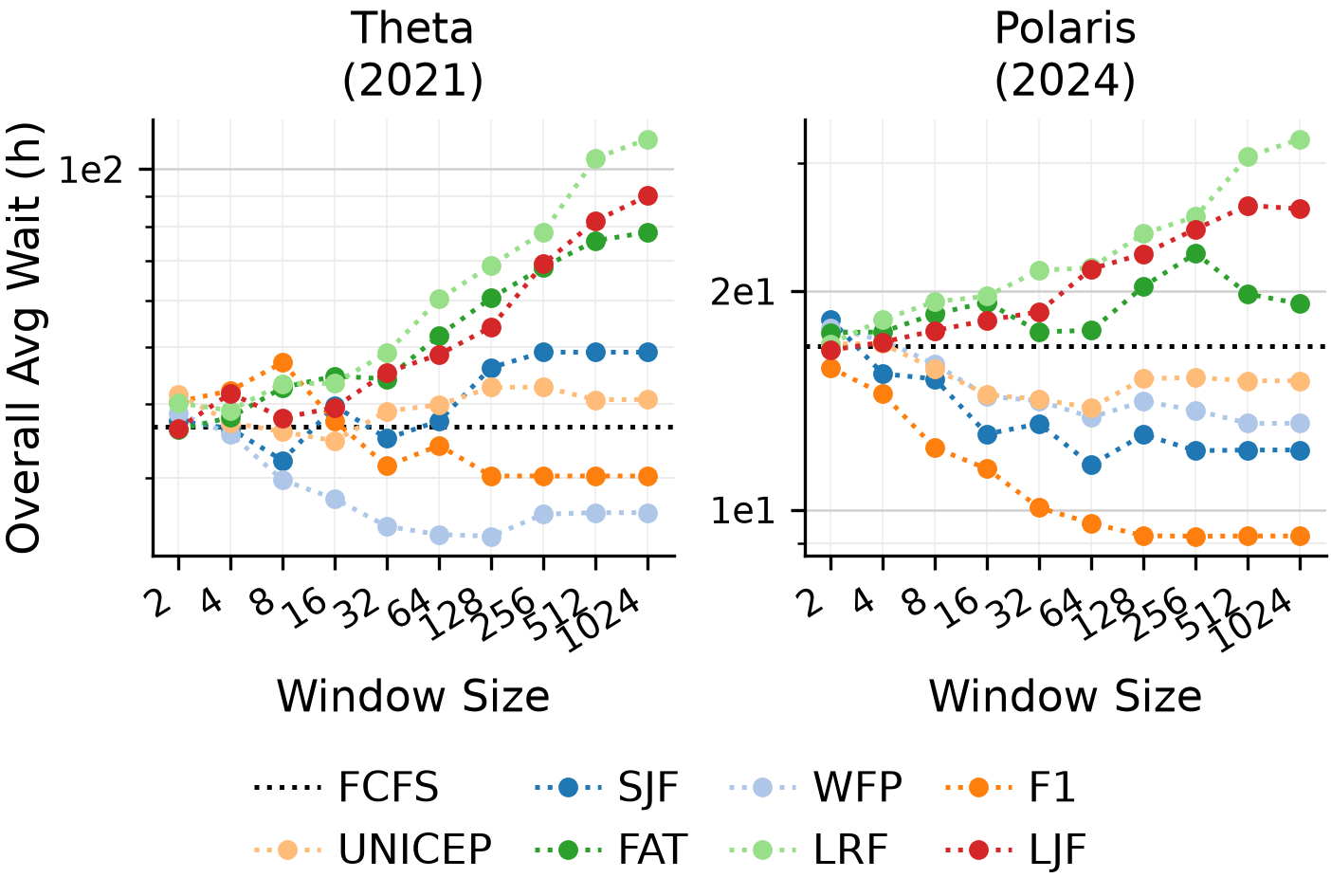}
    \caption{\small{Sensitivity of average wait time to window size w across heuristics on Theta and Polaris. Most heuristics achieve stable performance at $w=256$, whereas FAT, LJF, and LRF continue to degrade with larger window sizes. Unless otherwise noted, all evaluations use $w=256$.}}
    \label{fig:avg_wait_w}
\end{figure}

We compare MARS against common heuristic baselines and the state-of-the-art DRL scheduler RLScheduler, described below. As shown in Figure~\ref{fig:avg_wait_w}, $w=256$ minimizes average wait time across both systems and is therefore used for all heuristic baselines. All heuristics are paired with EASY backfilling~\cite{backfill}.

 \begin{itemize}
    \item \textit{WFP}: The production baseline on both Polaris and Theta. Paired with EASY backfilling at ALCF~\cite{allcock_expatanl_2018, tang-cluster09}, WFP scores jobs with $-(w_j / r_j)^3 \cdot (n_j)$, favoring older and shorter jobs while preventing starvation of large requests. This corresponds to WFP3 from Table \ref{tab:heuristics-policies}, where the $3$ is the power for the first term. Often referred to as the WFP in general.
    
    \item \textit{First-Come-First-Served (FCFS)}: The  widely used scheduling baseline; serves jobs strictly in arrival order.
        
    \item \textit{Shortest Job First (SJF)}: Prioritizes jobs by shortest requested walltime.
    
    \item \textit{F1}: Designed to minimize average bounded slowdown, F1 uses the scoring function in Table~\ref{tab:heuristics-policies}, derived via brute-force simulation and nonlinear regression. Its terms balance small, short jobs against starvation prevention for older requests.

     \item \textit{RLScheduler}: The DRL scheduler~\cite{zhang_rlscheduler_2020} is retrained to minimize bounded slowdown, which jointly weighs wait time and estimated runtime and serves as the optimization target in the original paper. We train the model on one year of 2020 Theta logs and monitor convergence via episodic returns; training requires approximately 8 hours on an NVIDIA A100 40GB GPU. The retrained model is evaluated on the 2021 Theta workload. To evaluate cross-system generalization, we test this model on the Polaris 2024 workload.
     
     \item \textit{Random}: While MCTS branches on 161 policies, many result in identical outcomes (Section \ref{sec:tree_pruning}). To evaluate the reward signal's impact on search guidance, we implement a random policy baseline that selects a root-level branch at random uniformly without performing further MCTS.
    \item \textit{MARS-CW \& MARS-CU}: Both reward configurations use a 15-second search time, consistent with standard HPC scheduling overhead constraints \cite{fan_deep_2021, fan_scheduling_2019}, and a tree depth ($D$) of 50. The exploration constant ($c$) is dynamically adjusted based on the range of $Q(s_t)$ values in the search tree, and a discount factor ($\gamma$) of 0.99 prioritizes long-term rewards. We use root parallelization \cite{parallel1} to construct 250 MCTS trees in parallel across 256 CPU cores within the search budget.
\end{itemize}

\section{Experimental Results}
\label{sec:exp_results}

Our experiments were conducted on a bare-metal node leased from Chameleon Cloud \cite{Chameleon}. The system was equipped with two 64-core AMD EPYC 7763 CPUs (128 physical cores and 256 hardware threads in total) and 256 GiB of RAM. The results  are structured into three parts. First, we analyze the wait times by examining overall performance and drawing conclusions based on backfill and drain behavior. Next, we assess utilization by examining the various uptimes and utilization for the 48 hours preceding downtime to evaluate behavior before maintenance reservations. Lastly, we evaluate the performance of our MARS agent using iterations achieved and the trend in the branching factor. 

\subsection{Job Wait Time}
\label{sec:wait_time}

\begin{figure}[t]
    \centering
    \includegraphics[width=0.9\linewidth]{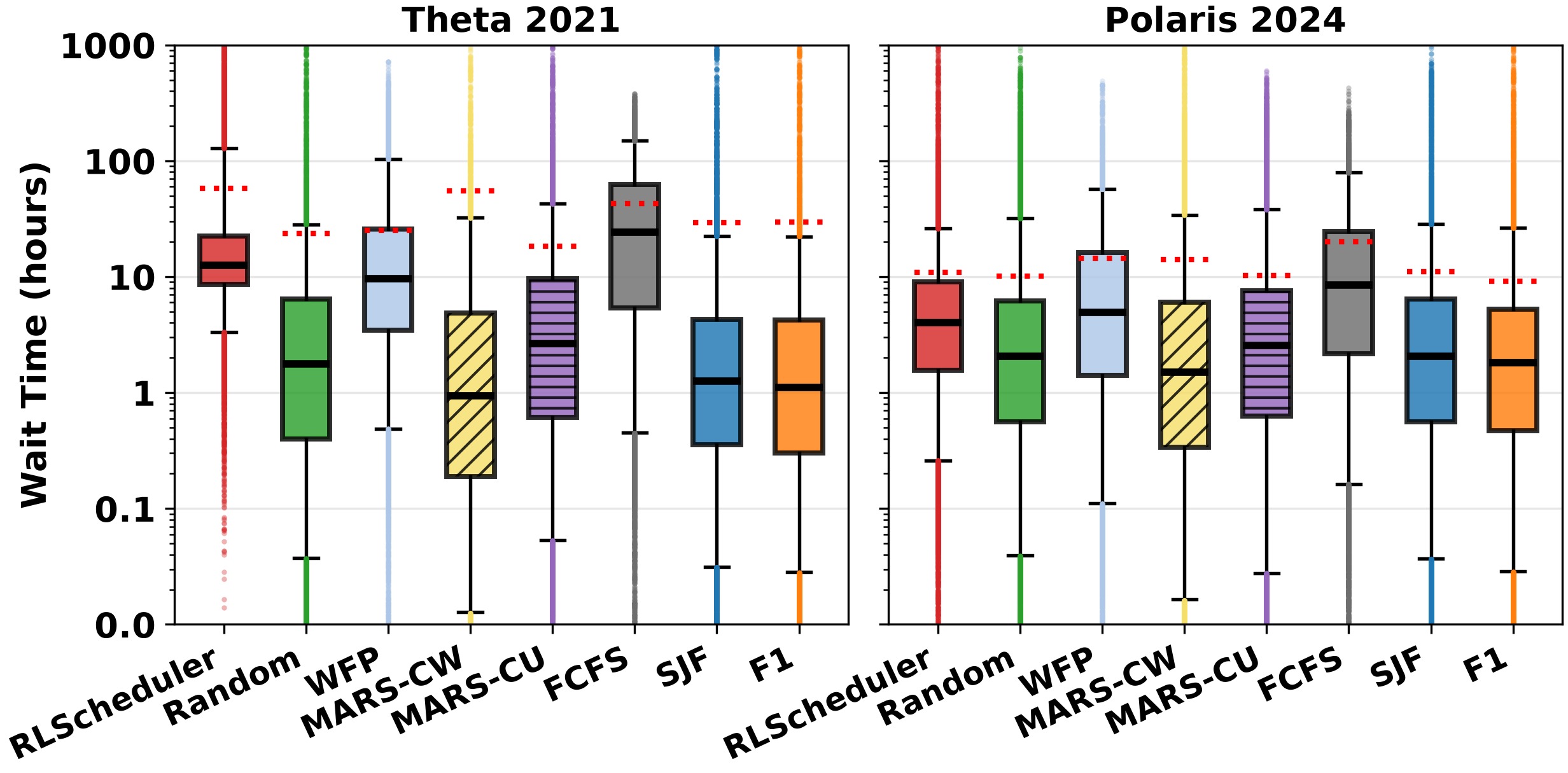}
    \caption{\small{Wait time distribution on Theta and Polaris. MARS-CW reduces $P_{75}$ wait time by 81\% on Theta and 55\% on Polaris relative to WFP (the production baseline).}}
    \label{fig:overall_wait}
\end{figure}

\subsubsection{Overall Performance}
Figure~\ref{fig:overall_wait} compares the overall wait time distribution across baselines and MARS configurations. First, RLScheduler exhibits the highest wait times on both systems, underperforming every heuristic and even the Random baseline; this is consistent with the convergence and generalization issues reported for pure DRL schedulers \cite{zhang_rlscheduler_2020}. Second, MARS-CW, which optimizes directly for wait time, achieves the lowest median and 75th-percentile wait times on both systems, with tail behavior comparable to the strongest heuristic baselines. Third, while SJF and F1 outperform the production baseline (WFP) on wait time, they are not used in production because they sacrifice system utilization, a trade-off quantified in Figure~\ref{fig:overall_util}. Notably, MARS-CW is not designed to outperform every heuristic on this single metric; rather, it provides competitive wait-time performance without retaining and providing flexibility to support production scheduling objectives through configurable reward design.

\subsubsection{Performance by Job Class}
\label{sec:results-jobsize}

\begin{figure*}[htbp]
  \centering
  \includegraphics[width=0.9\linewidth]{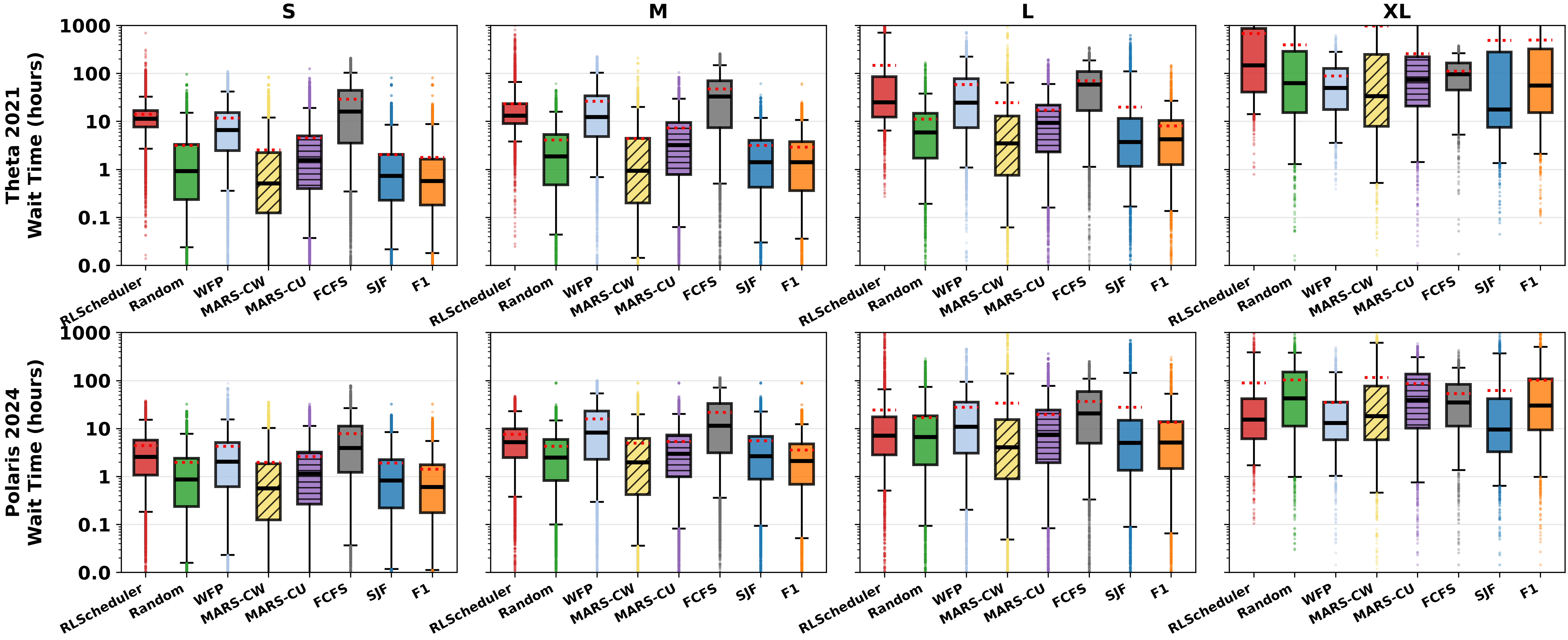}
  \caption{\small{Wait time distributions on Theta (top) and Polaris (bottom), broken down by job class (S, M, L, XL).  Compared to the production WFP baseline, MARS-CW reduces mean wait time for small and medium jobs by 78.9\% (S) and 83.2\% (M) on Theta, and by 48.2\% (S) and 63.8\% (M) on Polaris. 
  For larger jobs, the improvement concentrates in the median:  wait time drops by 83.1\% (L) and 32.1\% (L) on Polaris. The XL jobs mainly see degradation in performance.}}
  \label{fig:job_wise_wait}
\end{figure*}

Figure~\ref{fig:job_wise_wait} breaks down wait time by the size classes from Figure~\ref{fig:heatmap_workload}, revealing three patterns. First, nearly all policies exhibit increasing wait times as job size increases. Second, RLScheduler performs reasonably well for the S and M classes but degrades substantially for larger jobs, particularly on the larger Theta system. In contrast, MARS-CW achieves the largest reductions compared to the production baseline WFP in the mean wait time for the S and M classes, while also delivering substantial improvements in median wait time for the L and XL classes, whose jobs are most susceptible to starvation while waiting for contiguous resources. This behavior is also evident in the wait-time distributions shown in Figure~\ref{fig:drain_backfill_dist}, where MARS-CW exhibits flatter, more left-skewed distributions than WFP, indicating overall improvements despite some degradation for a small subset of jobs at the tail. Third, the Random baseline remains competitive with several heuristic schedulers, confirming that the heuristic action space itself is relatively strong. Nevertheless, both MARS-CW and MARS-CU outperform Random across nearly every job class, demonstrating that the learned reward signal effectively guides the scheduler toward its intended optimization objective.

\subsubsection{Draining \& Backfill Dynamics}
\label{sec:results-drainbf}

% \begin{figure}[htbp]
%     \centering
%     \includegraphics[width=\columnwidth]{figures/drain_overview_all_except_rlscheduler.png}
%     \caption{\small{Share of jobs scheduled via backfill or drain. \emph{Left:} overall rate across all jobs. \emph{Right:} breakdown by job class. \todo{This needs to be referenced in the text; if not, remove it or explain more in the caption here.}}}
%     \label{fig:drain_dist}
% \end{figure}

 \begin{figure}[h]
    \centering
    \includegraphics[width=0.8\columnwidth]{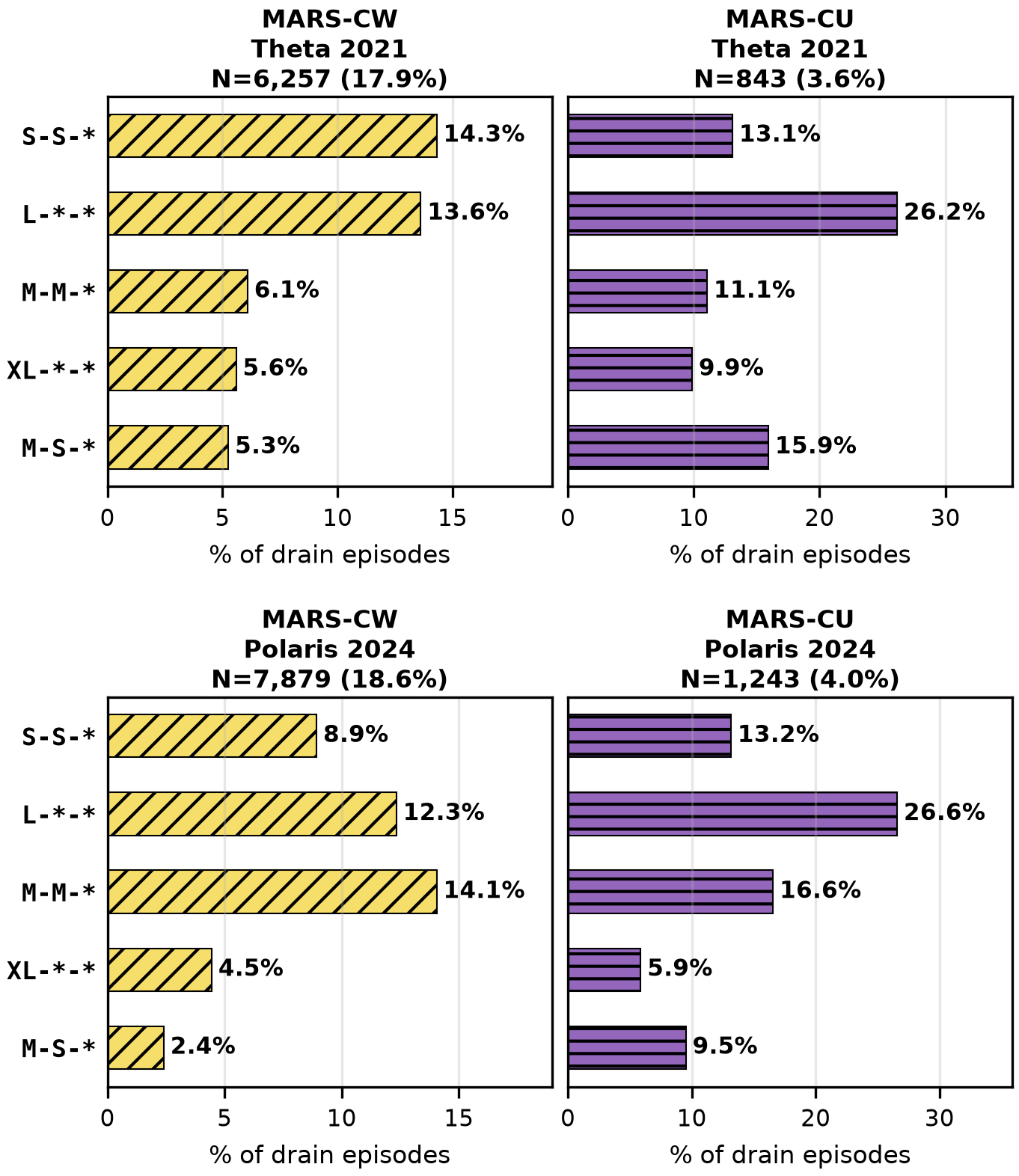}
    \caption{\small{Percentage of job sequences scheduled by MARS after draining. Each sequence reports the job class of the first two jobs, followed by a wildcard ($*$) for the third.
    (*) indicates a job class S/M/L/XL. Theta has a majority of S jobs, while Polaris has M jobs.}}
    \label{fig:drain_seq}
\end{figure}

 \begin{figure}[h]
    \centering
    \includegraphics[width=\columnwidth]{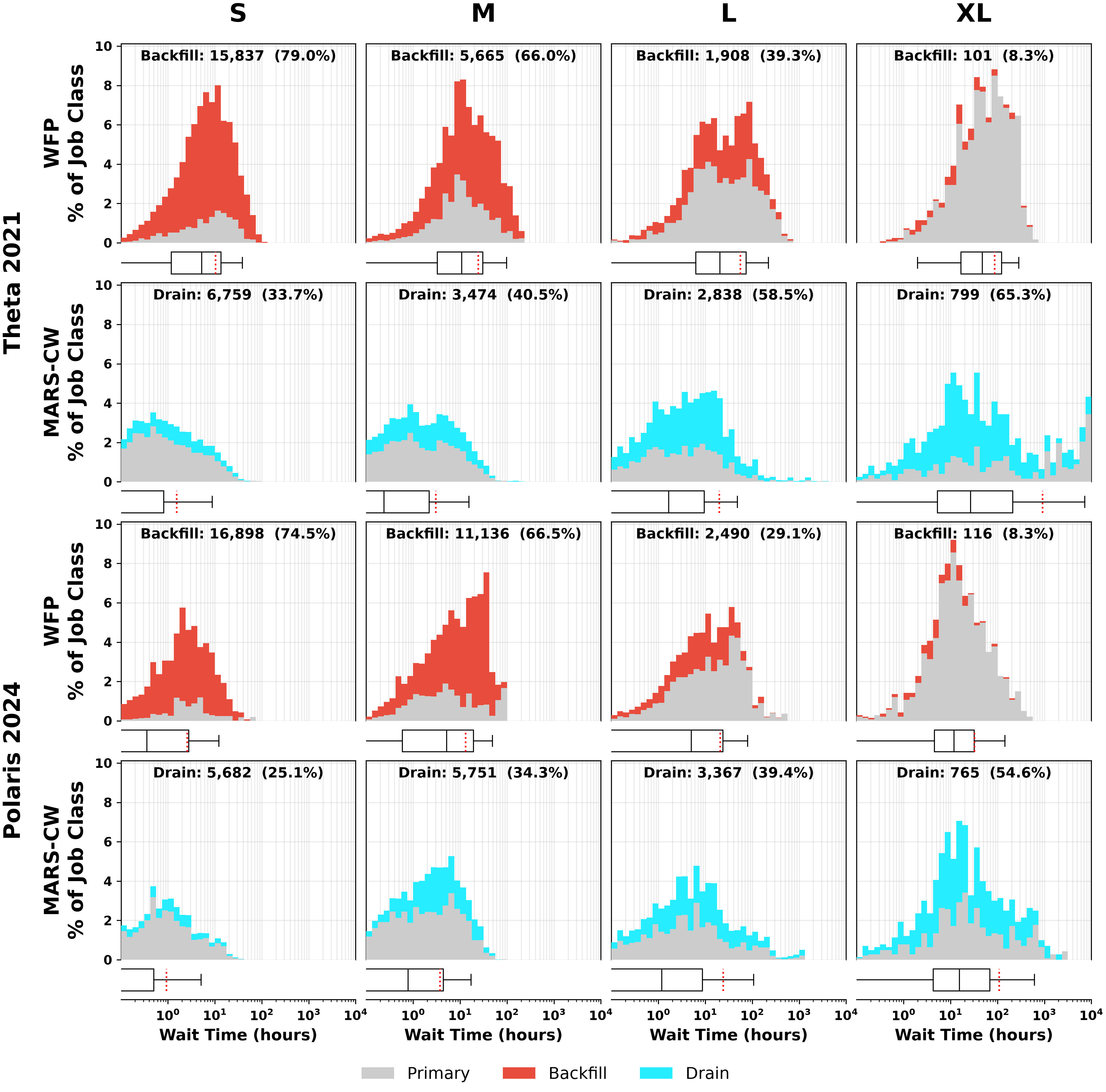}
    \caption{\small{Stacked wait time histograms by job class for the production baseline WFP (rows~1 and 3) and MARS-CW (rows~2 and 4) on Theta and Polaris. Bars are decomposed into Primary (gray), Backfill (red, WFP only), and Drain (cyan, MARS-CW only) contributions. A distribution skewed left indicates lower wait times.  MARS-CW clearly outperforms WFP in S/M/L but suffers at the tail for XL jobs}}
    \label{fig:drain_backfill_dist}
\end{figure}

The two key scheduling mechanisms examined in this study are traditional backfilling and the intelligent draining behavior that emerges from MARS. Draining occurs when MARS deliberately leaves available resources idle to create an allocation that better satisfies the optimization objective. Figure~\ref{fig:drain_backfill_dist} illustrates how conventional heuristics rely on backfilling compared to the draining behavior learned by MARS. For the production WFP scheduler, the effectiveness of backfilling decreases as job size increases, with larger jobs being backfilled less frequently. This behavior is expected, as backfilling primarily exploits small gaps in the schedule that are typically insufficient for large jobs. In contrast, the proportion of jobs scheduled after a drain increases with job size for MARS-CW, indicating that the scheduler intentionally creates contiguous resources for larger jobs. This behavior provides insight into the wait-time improvements for these job classes reported earlier.

The learned draining behavior also depends on the optimization objective. This is evident in Figure~\ref{fig:drain_seq}, where MARS-CW initiates drains more frequently than MARS-CU. This difference is expected because MARS-CU prioritizes high system utilization and therefore cannot afford to leave resources idle as often. Figure~\ref{fig:drain_seq} also categorizes the jobs scheduled immediately following a drain, showing that both reward configurations preferentially drain to schedule larger job classes. Interestingly, MARS-CW schedules S- and M-class jobs nearly as often after a drain as L- and XL-class jobs. This reflects the average wait-time objective, which naturally favors reducing the average waiting time across the entire workload. Since S and M jobs constitute the majority of the workload, scheduling them after a drain can produce a larger reduction in average wait time than exclusively prioritizing the largest jobs.

\subsection{System Utilization}
\label{sec:util}

\begin{figure}[htbp]
    \centering
    \includegraphics[width=0.9\linewidth]{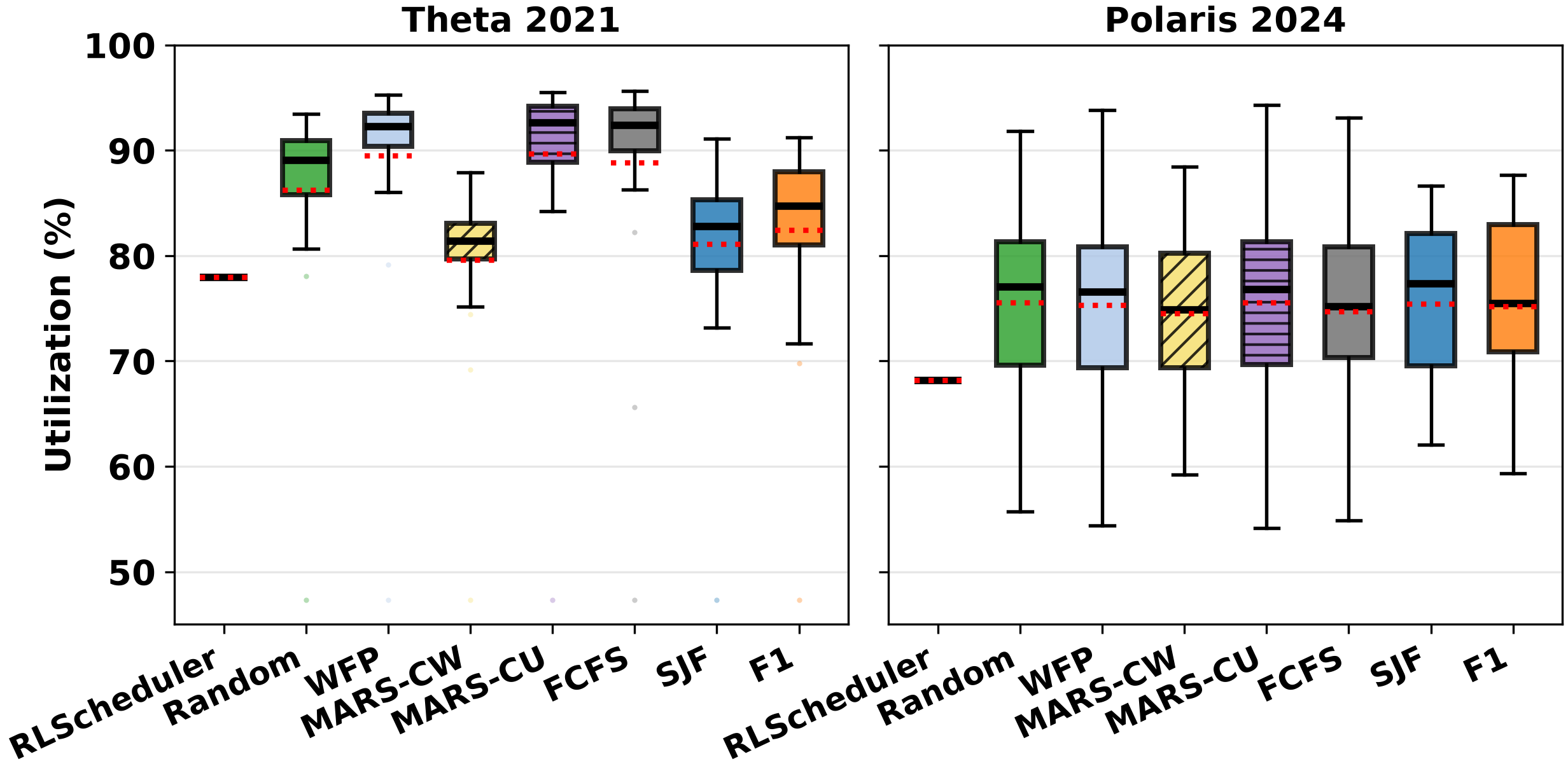}
    \caption{\small{Utilization distribution during system uptime. RLScheduler reports aggregate yearly utilization as it does not model downtime.}}
    \label{fig:overall_util}
\end{figure}

\begin{figure}[htbp]
  \centering
  \includegraphics[width=0.8\linewidth]{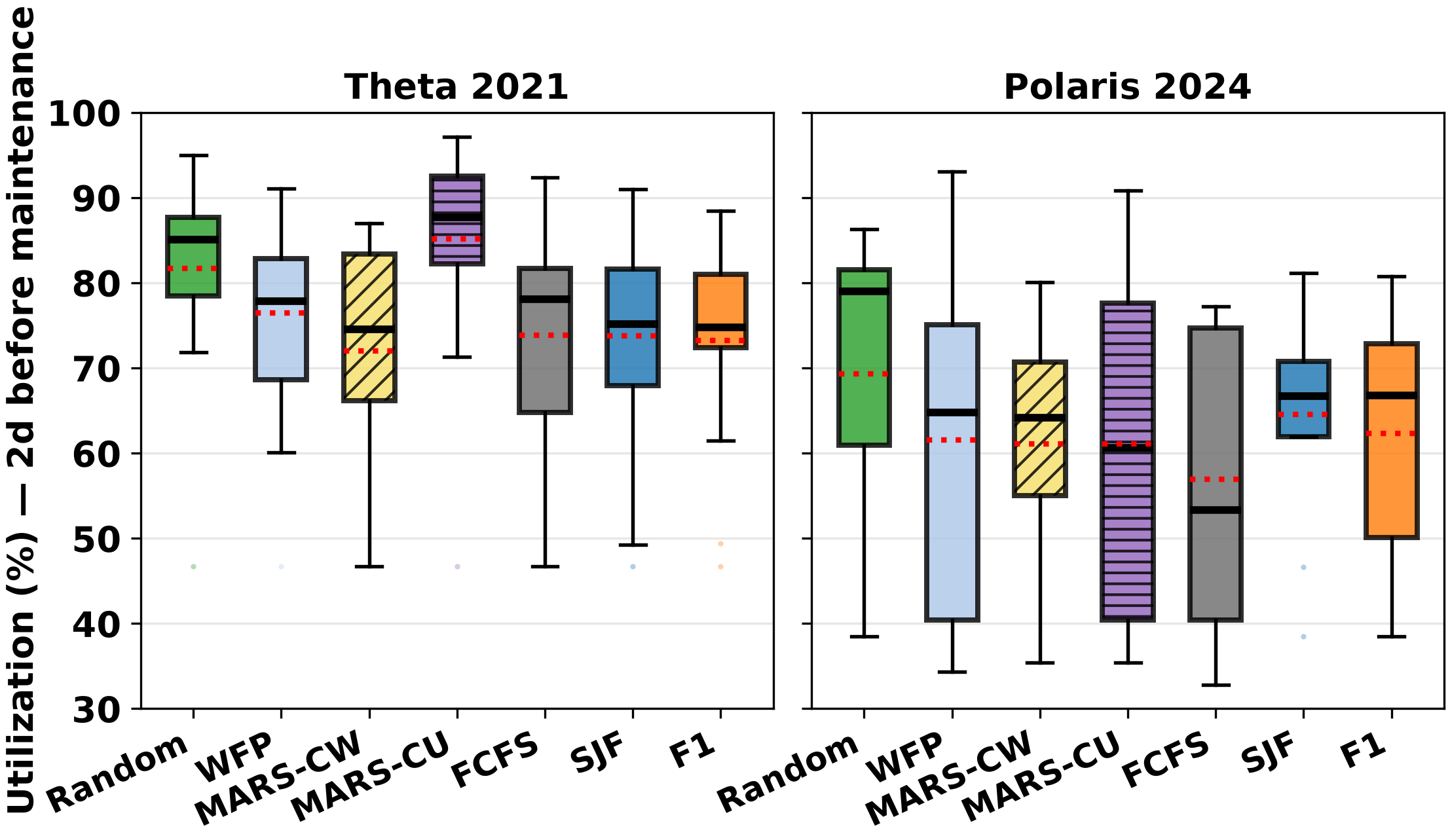}
   \caption{\small{Utilization distributions for the 48 hours preceding all scheduled downtimes. Theta had 21 downtimes, and Polaris had 9.}}
  \label{fig:drain_util}
\end{figure}

Figure \ref{fig:overall_util} illustrates the overall utilization for different uptime windows. MARS-CU performs exceptionally well on the Theta workload and slightly better on the Polaris workload. For RLScheduler, we report the utilization as a whole year since it lacks the modelling of downtimes. As mentioned in the previous section, MARS-CU effectively prevents system draining, thereby increasing its utilization. However, MARS-CW, F1, and SJF ultimately suffer due to the trade-off between optimizing for wait time and utilization. Specifically, MARS-CW loses out on utilization because of system draining.

Figure \ref{fig:drain_util} illustrates utilization over the 48 hours preceding scheduled downtime for both systems. On Theta, MARS-CU consistently outperforms the other methods across all utilization metrics. In contrast, Polaris exhibits minimal variance in utilization due to the nature of its workload: there are not enough jobs to fill the system’s capacity on the day before downtime. However, the heuristic baselines specifically show substantial variance before downtime on Theta, indicating instability in meeting utilization goals. MARS-CU, by contrast, maintains a more stable utilization level on Theta and achieves the utilization target in most cases.

\subsection{Search Throughput \& Branching}
\label{sec:mcts_runtime}

\begin{figure}[t]
  \centering
  \includegraphics[width=\linewidth]{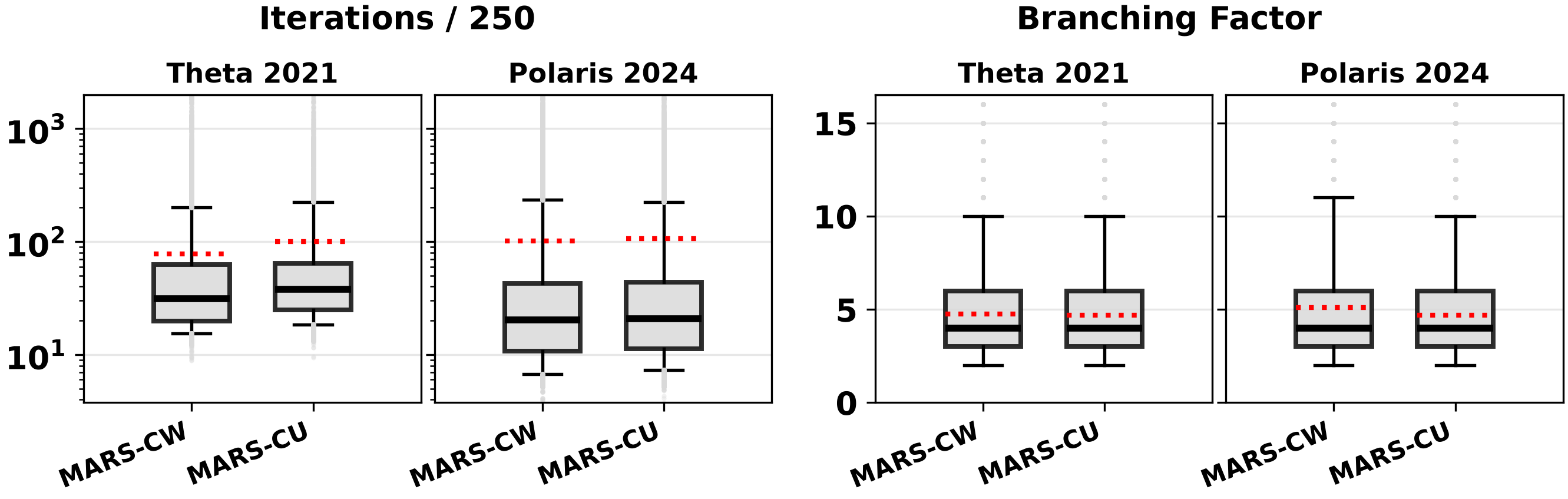} 
  \caption{\small{Per-cycle MCTS runtime characteristics on Theta and Polaris. \emph{Left:} distribution of MCTS iterations completed per scheduling cycle within the 15-second budget, normalized per parallel tree. \emph{Right:} effective branching factor at the root after node merging. }}
  \label{fig:mcts_perf}
\end{figure}

The number of iterations completed within the time budget is the standard measure of MCTS efficiency; more iterations mean more thorough exploration of the action space and, in principle, better decisions. Figure~\ref{fig:mcts_perf} characterizes both efficiency and search structure for MARS across the two systems.

The left panel shows that each of the 250 parallel trees completes a median of 30--40 iterations on Theta and 20 on Polaris within the 15-second scheduling budget, with upper whiskers past $10^3$. Aggregated across all trees, this yields thousands of iterations per scheduling decision, sufficient look-ahead depth for production workloads, enabled by the C++ parallel implementation.

The right panel reports the effective branching factor at the root after node merging. Although the theoretical action space contains 161 branches (16 heuristics $\times$ 10 window sizes plus FCFS), the effective branching factor sits at a median of 4-5 on both systems and rarely exceeds 10, confirming that resource constraints and redundant heuristic orderings collapse the action space substantially. MARS-CW and MARS-CU exhibit nearly identical branching distributions indicating that the reward signal shapes \emph{which} branch is selected, not \emph{how many} are available.

\section{Related Work}
\label{sec:related_work}

\paragraph{Heuristics}
The most prevalent scheduling approach in production systems relies on heuristic functions, such as First-Come-First-Served (FCFS) or sophisticated utility-based scoring \cite{backfill,allcock_expatanl_2018, tang-cluster09,alcfAuroraRunningJobs,nerscQueuesChargesPolicy}. These methods are transparent and do not require training, offering high stability across different scenarios. However, as noted in Table~\ref{tab:scheduling-comparison}, they lack the flexibility to adapt to changing optimization goals or complex constraints like machine reservations, as their scoring rules are rigid and myopic, not considering future states of the machine.

\paragraph{Deep Reinforcement Learning (DRL)} 
To improve upon static heuristics, DRL methods based on deep neural networks have been developed to learn scheduling policies through experience \cite{fan_deep_2021, zhang_rlscheduler_2020}. While they can outperform heuristics, they require extensive training and are highly sensitive to workload distributions. Furthermore, DRL approaches lack goal flexibility: changing optimization objectives typically requires retraining. They also struggle with cross-machine stability, as a policy trained on one system's workload often performs poorly when deployed on another with different characteristics \cite{zhang_rlscheduler_2020}.

\paragraph{Specialized Optimization}
This class of methods applies mathematical optimization techniques to address specific scenarios, such as adapting to shifting electricity prices \cite{yang-sc13} or maximizing burst buffer utilization \cite{fan_scheduling_2019}. These approaches support flexible objective design and do not require training, offering strong stability across systems.  However, they typically lack native support for dynamic machine reservations, which often requires additional mechanisms to incorporate look-ahead reasoning.

\paragraph{Innovation}
MCTS has been explored for combinatorial planning problems such as job shop scheduling \cite{jssp1, jssp2} and constrained staff scheduling \cite{staffsched}, where the underlying problems are NP-hard. 
\emph{As for HPC scheduling, this work is the first to use MCTS as a training-free, goal-configurable alternative to DRL}. 
Unlike DRL-based schedulers, which require extensive offline training and retraining when objectives change, our approach operates without any prior training and directly adapts to configurable goals at inference time. 
As shown in Table~\ref{tab:scheduling-comparison}, MCTS naturally handles machine reservations through explicit simulation of future scheduling states. It further provides high goal flexibility, allowing objectives such as wait time or utilization to be swapped without architectural modifications or expensive retraining.

\section{Conclusion}

We presented MARS, an MCTS-based HPC scheduler that requires no training and supports configurable optimization goals via a reward function. At each scheduling cycle, MARS adapts the search to the specified objective using the current system state within a fixed scheduling budget. Unlike DRL methods that require extensive training and often generalize poorly across systems, MARS produces decisions in seconds and scales via parallel tree construction.
Across two reward formulations, we show that distinct optimization goals can be achieved without model training, yielding more consistent, objective-specific performance than static heuristics, which necessarily trade off one metric against another. Overall, the results highlight two key properties of MARS: its ability to adapt search behavior to the configured reward without retraining, and its responsiveness to live system state within each scheduling cycle, proactively shaping the schedule rather than relying on backfill to create opportunities.

Three priorities guide our future work: refining MARS hyper-parameters through sensitivity studies (tree depth, budget); designing reward formulations that provide finer control over scheduling trade-offs; and developing scheduling models that address emerging energy-consumption challenges in HPC.

We have released MARS, along with the associated datasets and scheduling simulator, as open source on GitHub to support reproducibility and foster collaboration within the community \cite{MARS_GIT}.

\section*{Acknowledgment}
The authors wish to thank Professor Ian Kash (University of Illinois Chicago) for his insightful discussions. This work was supported in part by the U.S. National Science Foundation under Grants OAC 2402901 and CCF 2515009, and in part by the U.S. Department of Energy under Contract DE SC0024271.

\bibliographystyle{IEEEtran}
\bibliography{references}

\end{document}